\RequirePackage{fix-cm}
\documentclass[twocolumn,epjc3]{svjour3}
\smartqed  % flush right qed marks, e.g. at end of proof
\usepackage{amsfonts}
\usepackage{amsmath}
\usepackage{amssymb,bbold}
\usepackage{kantlipsum}
\RequirePackage{graphicx}
\RequirePackage{flushend}
\RequirePackage[colorlinks,citecolor=blue,urlcolor=blue,linkcolor=blue]{hyperref}
\RequirePackage[numbers,sort&compress]{natbib}
\RequirePackage{color}

 \RequirePackage{mathptmx}      % use Times fonts if available on your TeX system
\journalname{Eur. Phys. J. C}
\begin{document}

\title{Relativistic quantum dynamics of scalar bosons under a full vector Coulomb interaction%\thanksref{t1}
}


\author{Luis B. Castro\thanksref{e1,addr1} \and Luiz P. de Oliveira\thanksref{e2,addr2} \and 
Marcelo G. Garcia\thanksref{e4,addr4,addr5} \and Antonio S. de Castro\thanksref{e3,addr3}  %etc.
}
\thankstext{e1}{e-mail: lrb.castro@ufma.br, luis.castro@pq.cnpq.br}
\thankstext{e2}{e-mail: luizp@if.usp.br}
\thankstext{e4}{e-mail: marcelogarcia82@gmail.com}
\thankstext{e3}{e-mail: castro@pq.cnpq.br}

\institute{Departamento de F\'{\i}sica, Universidade Federal do Maranh\~{a}o (UFMA), Campus Universit\'{a}rio do Bacanga, 65080-805, S\~{a}o Lu\'{\i}s, MA, Brazil.
\label{addr1} \and
Instituto de F\'{\i}sica, Universidade de S\~{a}o Paulo (USP), 05508-900, S\~{a}o Paulo, SP, Brazil.\label{addr2} \and
Departamento de F\'{i}sica, Instituto Tecnol\'{o}gico de Aeron\'{a}utica (ITA), 12228-900, S\~{a}o Jos\'{e} dos Campos, SP, Brazil. \label{addr4} \and
Departamento de Matem\'{a}tica Aplicada, Universidade Estadual de Campinas (UNICAMP),
IMECC, 13081-970, Campinas, SP, Brazil.\label{addr5}
\and
Departamento de F\'{\i}sica e Qu\'{\i}mica, Universidade Estadual Paulista (UNESP), Campus de Guaratin\-gue\-t\'{a}, 12516-410, Guaratinguet\'{a}, SP, Brazil.\label{addr3}
}

\date{Received: date / Accepted: date}

\maketitle

\begin{abstract}
The relativistic quantum dynamics of scalar bosons in the background of a
full vector coupling (minimal plus nonminimal vector couplings) is explored
in the context of the Duffin-Kemmer-Petiau formalism. The Coulomb phase shift is determined for a general mixing of couplings and it is shown that the space component of the nonminimal coupling is a {\it sine qua non} condition for the exact closed-form scattering amplitude. It follows that the Rutherford cross section vanishes in the absence of the time component of the minimal coupling. Bound-state solutions obtained from the poles of the partial scattering amplitude show that the time component of the minimal coupling plays an essential role. The bound-state solutions depend on the nonminimal coupling and the spectrum consists of particles or antiparticles depending on the sign of the time component of the minimal coupling without chance for pair production even in the presence of strong couplings. It is also shown that an accidental degeneracy appears for a particular mixing of couplings.

\PACS{03.65.Pm \and 03.65.Ge}
\end{abstract}

\section{Introduction}
\label{intro}

The sucess of the Dirac equation in describing proton-nucleus scattering encourages the use of others fundamental relativistic wave equations in treating other nuclear probes. In the 1930's, R. Duffin, N. Kemmer and G. Petiau proposed a new
relativistic wave equation able to describe the dynamics of spin-zero and
spin-one particles \cite{Petiau1936,Kemmer1938,PR54:1114:1938,Kemmer1939}. The Duffin--Kemmer--Petiau (DKP) formalism holds a large number of couplings, not only electromagnetic interactions, that enables emulate a mean field theory for describing hadron interactions. This large number of couplings makes the DKP equation a great tool for physiscists that use a phenomenological description of nuclear processes. The DKP formalism has been widely
used in the description of many processes in elementary particle and nuclear
physics as for instance, in the
analysis of $K_{l3}$ decays, the decay-rate ratio $\Gamma(\eta\rightarrow
\gamma\gamma)/\Gamma(\pi^{0}\rightarrow \gamma\gamma)$, and level shifts and
widths in pionic atoms \cite{PRL26:1200:1971,PTP51:1585:1974,PRC34:2244:1986}%
. Kozac and collaborators \cite{PRC37:2898:1988} found that the DKP-based
deuteron-nucleus optical potential are in close agreement with those
obtained in other approachs \cite{OSAKA1985,NP8:484:1958}. Fischbach and
co-authors definite a testable prediction of a kinematic zero at $%
t=(m_{K}+m_{\pi })^{2}$ in the ``effective" scalar form factor \cite{PRL26:1200:1971}.
The same authors conclude that, based in others studies \cite{PRD6:726:1972,PRD7:207:1973}, the DKP equation is superior to the Klein-Gordon (KG) equation for the
description of scalar particles \cite{PRD7:3544:1973}. Willey and collaborators 
\cite{PRD7:3540:1973} applied the standard $S-$matrix kinematic analysis and showed that
the form factors are free of kinematic singularities or constraints. Aydin and Barut \cite{PRD7:3522:1973} obtained the energy spectra and the
branching ratio of the two $K_{l3}$ modes and the $\pi _{e3}$ decay rate
with good agreement to experiment. The DKP formalism has also applications in other contexts, as such, in
noncommutative phase space \cite{EPJC72:2217:2012}, in Very Special
Relativity (VSR) symmetries \cite{EPJP129:246:2014}, in Bose-Einstein
condensates \cite{PLA316:33:2003,PA419:612:2015}, in topological defects
\cite{EPJC75:287:2015,EPJP130:140:2015}, in thermodynamics properties \cite%
{ADHEP2015:901675:2015}, in topological semimetals \cite{PRB92:235106:2015}, in noninertial effect of rotating frames \cite{EPJC76:61:2016}, among others.

Although the formalisms are equivalent in the case of minimally coupled
vector interactions \cite{PLA244:329:1998,PLA268:165:2000,PRA90:022101:2014}%
, the DKP formalism enjoys a richness of couplings not capable of being
expressed in the KG and Proca theories \cite{PRD15:1518:1977,JPA12:665:1979}%
. The nonminimal vector interaction refers to a kind of charge conjugate
invariant coupling that behaves like a vector under a Lorentz
transformation. Nonminimal vector potentials, added by other kinds of
Lorentz structures, have already been used in a phenomenological context for
describing the scattering of mesons by nuclei \cite%
{PRL55:592:1985,PRC34:2240:1986,NPA585:311:1995,PLB427:231:1998}, but it
should be mentioned that the nonminimal vector couplings have been
improperly used. Other misconception is found in Refs. \cite%
{PRL55:592:1985,PRC34:2240:1986}, where the space component of the
nonminimal vector potential is absorbed into the spinor. However, as it is
shown in \cite{JPA43:055306:2010}, there is no chance to discard this term.
Recently, the demand for a conserved four-current has been used to point out
a misleading treatment in the literature regarding analytical solutions for
nonminimal vector interactions (see \cite{ADHEP2014:784072:2014} for a
comprehensive list of references). Elsewhere \cite{PLA375:2596:2011}, using
the proper form of the nonminimal vector coupling and considering
spherically symmetric potential functions, it has been shown that the
solution for the problem can be found in a clear and transparent way in
terms of a Schr\"{o}dinger-like equation for just one component of the DKP
spinor and the remaining components are expressed in terms of that one in a
simple way. Also, it has been shown that the proper boundary conditions are
imposed in a simple way by observing the absence of Dirac delta potentials.
On the other hand, the Coulomb problem has been intensively studied because
of its intrinsic interest and also for applications in different research
fields. The Coulomb problem in the context of the DKP equation has been
reported in the literature for minimal vector Coulomb interaction \cite%
{JPG19:87:1993,JMP35:4517:1994,PLB574:197:2003} and for scalar Coulomb
interaction \cite{PRC84:064003:2011}. To the best of our knowledge, no one
has reported on the solution of scalar bosons with a full vector Coulomb
interaction.

In the present work, we show the correct use of the nonminimal vector
interaction in view of misconceptions propagated in the literature and we
address the problem of scalar bosons embedded in a full vector Coulomb
potential. We show that the Coulomb interaction (whether attractive or
repulsive) leads to a Whittaker differential equation. The behavior of the
scattering and bound-state solutions as well as the restrictions on the
potential parameters are discussed in detail. This work treats mathematical aspects that may be important in nuclear physics but are important by themselves because they are closed analytical results. Mesons are submitted to strong interactions that are emulated on the mean field theory by a external potential. Beyond its intrinsic interest, the Coulomb interaction is a long--range like, but allows us to obtain analytical solutions that can give us a heuristic look of strong interactions in the tail, so that the new results reported in the present work are very important for a
better understanding in the phenomenological description of elastic
meson-nucleus scattering.

This work is organized as follows. In section \ref{sec:1}, we give a brief
review on the DKP equation. We discuss the condition on the interactions
which leads to a conserved current (section \ref{subsec:1:1}). In section %
\ref{sec:2}, we concentrate our efforts in the full vector interaction. In
particular, we focus the case of scalar bosons and obtain the equation of
motion, which describes the relativistic quantum dynamics (section \ref%
{subsec:2:1}). Considering the Coulomb interaction, we find the scattering
and bound-state solutions (section \ref{subsec:2:3} and \ref{subsec:2:4},
respectively). Finally, in section \ref{sec:3} we present our conclusions.

\section{Review on DKP equation}
\label{sec:1}

We consider the Lagrangian density for the free DKP field (with units in which $\hbar =c=1$)
\begin{equation}\label{lagran}
\mathcal{L}=\frac{i}{2}\left[ \bar{\psi}\beta^{\mu}\left(\partial_{\mu}\psi\right)-\left(\partial_{\mu}\bar{\psi}\right)\beta^{\mu}\psi \right]-m\bar{\psi}\psi
\end{equation}
\noindent where the adjoint spinor $\bar{\psi}$ is given by $\bar{\psi}=\psi
^{\dagger }\eta ^{0}$ with $\eta ^{0}=2\beta ^{0}\beta ^{0}-1$ in such a way
that $(\eta^{0}\beta^{\mu})^{\dag}=\eta^{0}\beta^{\mu}$ (the matrices $%
\beta^{\mu}$ are Hermitian with respect to $\eta^{0}$). The matrices $\beta ^{\mu }$\ satisfy the algebra
\begin{equation}  \label{betaalge}
\beta^{\mu }\beta ^{\nu }\beta ^{\lambda }+\beta ^{\lambda }\beta ^{\nu
}\beta ^{\mu }=g^{\mu \nu }\beta ^{\lambda }+g^{\lambda \nu }\beta ^{\mu }
\end{equation}
\noindent and the metric tensor is $g^{\mu \nu }=\,$diag$\,(1,-1,-1,-1)$.
The algebra expressed by (\ref{betaalge}) generates a set of 126 independent
matrices whose irreducible representations are a trivial representation, a
five-dimensional representation describing the spin-zero particles and a
ten-dimensional representation associated to spin-one particles.

The equation of motion obtained from the Lagrangian (\ref{lagran}), is given by \cite{Kemmer1939}
\begin{equation}
\left( i\beta ^{\mu }\partial _{\mu }-m\right) \psi =0  \label{dkp}
\end{equation}%
\noindent This equation is covariant under Lorentz transformation (see \ref{A:1})

A well-known conserved four-current is given by
\begin{equation}  \label{cu}
J^{\mu }=\frac{1}{2}\,\bar{\psi}\beta ^{\mu }\psi
\end{equation}
\noindent where the adjoint spinor $\bar{\psi}$ is given by $\bar{\psi}=\psi
^{\dagger }\eta ^{0}$ with $\eta ^{0}=2\beta ^{0}\beta ^{0}-1$ in such a way
that $(\eta^{0}\beta^{\mu})^{\dag}=\eta^{0}\beta^{\mu}$ (the matrices $%
\beta^{\mu}$ are Hermitian with respect to $\eta^{0}$). Despite the
similarity to the Dirac equation, the DKP equation involves singular
matrices, the time component of $J^{\mu}$ is not positive definite and the
case of massless bosons cannot be obtained by a limiting process~\cite%
{PRD10:4049:1974}. Nevertheless, the matrices $\beta^{\mu}$ plus the unit
operator generate a ring consistent with integer-spin algebra and $J^{0}$
may be interpreted as a charge density. The factor $1/2$ multiplying $\bar{%
\psi}\beta^{\mu}\psi$, of no importance regarding the conservation law, is
in order to hand over a charge density conformable to that one used in the
KG theory and its nonrelativistic limit \cite{JPA43:055306:2010}. The
normalization condition for bound-state solutions $\int d\tau \,J^{0}=\pm 1$
can be expressed as%
\begin{equation}
\int d\tau \,\bar{\psi}\beta ^{0}\psi =\pm 2  \label{norm}
\end{equation}
\noindent where the plus (minus) sign must be used for a positive (negative)
charge.

\subsection{Interaction in the Duffin-Kemmer-Petiau equation}

\label{subsec:1:1}

With the introduction of interactions, the Lagrangian density for the DKP field becomes%
\begin{equation}\label{lagran2}
\mathcal{L}=\frac{i}{2}\left[ \bar{\psi}\beta^{\mu}\left(\partial_{\mu}\psi\right)-\left(\partial_{\mu}\bar{\psi}\right)\beta^{\mu}\psi \right]-m\bar{\psi}\psi-\bar{\psi}U\psi
\end{equation}
\noindent where the more general potential matrix $U$ is written in terms of
25 (100) linearly independent matrices pertinent to five (ten)-dimensional
irreducible representation associated to the scalar (vector) sector. The equation of motion obtained from the Lagrangian (\ref{lagran2}), is given by
\begin{equation}
\left( i\beta ^{\mu }\partial _{\mu }-m-U\right) \psi =0\,.  \label{dkp2}
\end{equation}%
\noindent In the presence of interaction, $J^{\mu }$ satisfies the equation
\begin{equation}
\partial _{\mu }J^{\mu }+\frac{i}{2}\,\bar{\psi}\left( U-\eta ^{0}U^{\dagger
}\eta ^{0}\right) \psi =0\,.  \label{corrent2}
\end{equation}%
\noindent Thus, if $U$ is Hermitian with respect to $\eta ^{0}$ then
four-current will be conserved. The potential matrix $U$ can be written in
terms of well-defined Lorentz structures. For the spin-zero sector there are
two scalar, two vector and two tensor terms \cite{PRD15:1518:1977}, whereas
for the spin-one sector there are two scalar, two vector, a pseudoscalar,
two pseudovector and eight tensor terms \cite{JPA12:665:1979}. The tensor
terms have been avoided in applications because they furnish noncausal
effects \cite{PRD15:1518:1977,JPA12:665:1979}. The condition (\ref{corrent2}%
) has been used to point out a misleading treatment in the recent literature
regarding solutions for nonminimal vector interactions \cite%
{ADHEP2014:784072:2014,CJP87:857:2009,CJP87:1185:2009,JPA45:075302:2012}.

\section{Vector interactions in the DKP equation}

\label{sec:2}

Considering only the vector terms, the DKP equation can be written as
\begin{equation}
\left( i\beta ^{\mu }\partial _{\mu }-m-\beta^{\mu}A_{\mu}^{(1)}-i[P,\beta
^{\mu }]A_{\mu }^{(2)}\right) \psi =0  \label{dkp02}
\end{equation}
\noindent where $P$ is a projection operator ($P^{2}=P$ and $P^{\dagger }=P$%
) in such a way that $\bar{\psi}[P,\beta ^{\mu }]\psi $ behaves like a
vector under a Lorentz transformation as does $\bar{\psi}\beta ^{\mu }\psi $%
. Here, $A_{\mu}^{(1)}$ and $A_{\mu}^{(2)}$ are the four--vector potential
functions. Notice that the vector potential $A_{\mu}^{(1)}$ is minimally
coupled but not $A_{\mu}^{(2)}$. One very important point to note is that
this potential leads to a conserved four-current but the same does not
happen if instead of $i[P,\beta^{\mu}]$ one uses either $P\beta^{\mu}$ or $%
\beta^{\mu}P$, as in \cite%
{PRL55:592:1985,PRC34:2240:1986,NPA585:311:1995,PLB427:231:1998}. As a
matter of fact, in \cite{PRL55:592:1985} it is mentioned that $P\beta^{\mu}$
and $\beta^{\mu}P$ produce identical results.

The invariance of the nonminimal vector potential under charge conjugation
means that it does not couple to the charge of the boson. In other words, $%
A_{\mu}^{(2)}$ does not distinguish particles from antiparticles. Hence,
whether one considers spin-zero or spin-one bosons, this sort of interaction
cannot exhibit Klein's paradox~\cite{JPA43:055306:2010}. On the other hand,
the charge--conjugation operation changes the sign of the minimal
interaction potential, i.e changes the sign of $A_{\mu}^{(1)}$ (see \ref{A:2}).

If the potential is time-independent one can write $\psi (\vec{r},t)=\phi (%
\vec{r})\exp (-iEt)$, where $E$ is the energy of the boson, in such a way
that the time-independent DKP equation becomes%
\begin{equation}
\left( \beta ^{0}E+i\vec{\beta}\cdot\vec{\nabla}-m-\beta^{\mu}A_{%
\mu}^{(1)}-i[P,\beta ^{\mu }]A_{\mu }^{(2)} \right) \phi =0\,.  \label{DKP10}
\end{equation}%
\noindent In this case \ $J^{\mu }=\bar{\phi}\beta ^{\mu }\phi /2$ does not
depend on time, so that the spinor $\phi$ describes a stationary state.

\subsection{Scalar sector}

\label{subsec:2:1}

For the case of spin-zero (scalar sector), the $\beta ^{\mu }$\ matrices are~%
\cite{JPA45:075302:2012}%
\begin{equation}
\beta ^{0}=%
\begin{pmatrix}
\theta & \overline{0} \\
\overline{0}^{T} & \mathbf{0}%
\end{pmatrix}%
,\quad \vec{\beta}=%
\begin{pmatrix}
\widetilde{0} & \vec{\rho } \\
-\vec{\rho }^{\,T} & \mathbf{0}%
\end{pmatrix}%
\end{equation}%
\noindent where%
\begin{eqnarray}
\ \theta &=&%
\begin{pmatrix}
0 & 1 \\
1 & 0%
\end{pmatrix}%
,\quad \rho ^{1}=%
\begin{pmatrix}
-1 & 0 & 0 \\
0 & 0 & 0%
\end{pmatrix}
\notag \\
&& \\
\rho ^{2} &=&%
\begin{pmatrix}
0 & -1 & 0 \\
0 & 0 & 0%
\end{pmatrix}%
,\quad \rho ^{3}=%
\begin{pmatrix}
0 & 0 & -1 \\
0 & 0 & 0%
\end{pmatrix}
\notag
\end{eqnarray}
\noindent $\overline{0}$, $\widetilde{0}$ and $\mathbf{0}$ are 2$\times $3, 2%
$\times $2 and 3$\times $3 zero matrices, respectively, while the
superscript T designates matrix transposition. In this representation $P=%
\frac{1}{3}\left(\beta^{\mu }\beta _{\mu }-1\right)=\mathrm{diag}%
\,(1,0,0,0,0)$, i.e. $P$ projects out the first component of the DKP spinor.
The five-component spinor can be written as $\phi ^{T}=\left( \phi
_{1},...,\phi _{5}\right)$ and the three-dimensional DKP equation for scalar
bosons becomes
\begin{equation}\label{dkp11}
\left[ \left(i\vec{\nabla}+\vec{A}^{(1)} \right)^{2}+\vec{\nabla }\cdot \vec{%
A}^{(2)}+\left(\vec{A}^{(2)}\right)^{2}\right] \phi _{1}=k^{2}\phi _{1}
\end{equation}%
\begin{equation}\label{dkp3a}
\phi _{2}=\frac{1}{m}\left( E-A_{0}^{(1)}+iA_{0}^{(2)}\right) \,\phi _{1}
\end{equation}
\begin{equation}\label{dkp3}
\vec{\zeta }=\left( \vec{\nabla }-i\vec{A}^{(1)}-\vec{A}^{(2)}\right)
\phi_{1} 
\end{equation}
\noindent where%
\begin{equation}
k^{2}=\left(E-A_{0}^{(1)}\right)^{2}-m^{2}+\left(A_{0}^{(2)}\right)^{2}
\label{k}
\end{equation}
\noindent and
\begin{equation}
\vec{\zeta }=\left( \zeta_{1},\zeta_{2},\zeta_{3} \right)=\frac{m}{i}(\phi
_{3},\phi _{4},\phi _{5})\,.
\end{equation}
\noindent Note that the equations (\ref{dkp11}), (\ref{dkp3a}) and (\ref{dkp3}) clearly show that the DKP spinor has an excess of components. We can see that $\phi_{1}$ is the independent component of the DKP spinor $\phi$, which is associated to correct physical component and represents a complex scalar field \cite{PRA90:022101:2014}. The other $4$ components of $\phi$ are superfluous components, which are expressed in terms of $\phi_{1}$.

\noindent Meanwhile,
\begin{equation}
\begin{split}
J^{0}=\frac{E-A_{0}^{(1)}}{m}\,|\phi _{1}|^{2}\,, \\
\vec{J}=\frac{1}{m}\left[\mathrm{Im}\left( \phi _{1}^{\ast }\,\vec{\nabla }%
\phi _{1}\right)- \vec{A}^{(1)}|\phi _{1}|^{2} \right]\,.
\end{split}%
\end{equation}%
\noindent

If we consider spherically symmetric potentials
\begin{equation}
\left( A^{(\alpha )}\right) ^{\mu }(\vec{r})=\left( A_{0}^{(\alpha
)}(r),A_{r}^{(\alpha )}(r)\vec{\widehat{r}}\right) \,,\quad \alpha =1,2
\label{pote}
\end{equation}%
\noindent then the DKP equation permits the factorization
\begin{equation}
\phi _{1}(\vec{r})=\frac{u_{\kappa }(r)e^{i\Lambda }}{r}\,Y_{lm_{l}}(\theta
,\varphi )  \label{sp}
\end{equation}%
\noindent where $A_{r}^{(1)}=d\Lambda /dr$, $Y_{lm_{l}}$ is the usual
spherical harmonic, with $l=0,1,2,\ldots $, $m_{l}=-l,-l+1,\ldots ,l$, $\int
d\Omega \,Y_{lm_{l}}^{\ast }Y_{l^{\prime }m_{l^{\prime }}}=\delta
_{ll^{\prime }}\delta _{m_{l}m_{l^{\prime }}}$ and $\kappa $ stands for all
quantum numbers which may be necessary to characterize $\phi _{1}$. For $%
r\neq 0$ the radial function $u$ obeys the radial equation
\begin{equation}\label{21}
\frac{d^{2}u}{dr^{2}}+\left[ k^{2}-2\,\frac{A_{r}^{(2)}}{r}-\frac{%
dA_{r}^{(2)}}{dr}-\frac{l(l+1)}{r^{2}}-\left( A_{r}^{(2)}\right) ^{2}\right]
u=0 \,. 
\end{equation}%
\noindent If the potentials $A_{0}^{(1)}$, $A_{0}^{(2)}$ and $A_{r}^{(2)}$
go to zero at large distances the proper solution has the asymptotic
behavior $e^{iKr}$ as $r\rightarrow \infty $, with
\begin{equation}
K=\sqrt{E^{2}-m^{2}}\,.  \label{K}
\end{equation}%
\noindent Therefore, scattering states occur only if $K\in \mathbb{R}$,
whereas bound states occur only if $K=i|K|$. Furthermore, in the case of
bound-state solutions the condition $\int d\tau \,J^{0}=\pm 1$
implies%
\begin{equation}
\frac{E}{m}\int_{0}^{\infty }dr\,|u|^{2}-\frac{1}{m}\int_{0}^{\infty
}dr\,A_{0}^{(1)}|u|^{2}\,=\pm 1\,.  \label{corre}
\end{equation}%
\noindent Therefore, for motion in a central field, the solution of the
three-dimensional DKP equation can be found by solving a Schr\"{o}%
dinger-like equation. The other components are obtained from (\ref{dkp3a})
and (\ref{dkp3}). Due to the spherical symmetry, Eq. (\ref{21}) does not
depend on $m_{l}$. Hence, for each $l$ the energy is degenerate $2l+1$ times
(essential degeneracy).

\subsection{Full vector Coulomb potential}

\label{subsec:2:2}

Let us consider the vector terms in the form
\begin{equation}
A_{0}^{(1)}=\frac{a_{0}^{(1)}}{r},\qquad A_{0}^{(2)}=\frac{a_{0}^{(2)}}{r}%
,\qquad A_{r}^{(2)}=\frac{a_{r}^{(2)}}{r}\,.  \label{potl}
\end{equation}%
\noindent Substituting (\ref{potl}) in (\ref{21}), we obtain
\begin{equation}  \label{eqcou1}
\frac{d^{2}u}{dr^{2}}+\left[ K^{2}-\frac{\alpha _{1}}{r}-\frac{\alpha
_{2}+l(l+1)}{r^{2}}\right] u=0 \,,
\end{equation}%
\noindent with
\begin{eqnarray}
\alpha _{1} &=&2Ea_{0}^{(1)}\,,  \label{lambdaa} \\
\alpha _{2}
&=&a_{r}^{(2)}(a_{r}^{(2)}+1)-\left(a_{0}^{(1)}\right)^{2}-\left(a_{0}^{(2)}%
\right)^{2} \,.  \label{etaa}
\end{eqnarray}%
\noindent Using the abbreviations
\begin{equation}  \label{gammal}
\gamma_{l}=\sqrt{\left( l+\frac{1}{2} \right)^{2}+\alpha_{2}}\,,
\end{equation}
\begin{equation}  \label{eta}
\eta=\frac{ \alpha_{1}}{2K}\,,
\end{equation}
\noindent and the change $z=-2iKr$, the equation (\ref{eqcou1}) becomes
\begin{equation}  \label{whit}
\frac{d^{2}u}{dz^{2}}+\left( -\frac{1}{4}-\frac{i\eta}{z}+\frac{%
1/4-\gamma_{l}^{2}}{z^{2}} \right)u=0\,.
\end{equation}
\noindent This second-order differential equation is the called Whittaker
equation, which have two linearly independent solutions $M_{-i\eta,%
\gamma_{l}}(z)$ and $W_{-i\eta,\gamma_{l}}$ behaving like $z^{1/2+\gamma_{l}}
$ and $z^{1/2-\gamma_{l}}$ close to the origin, respectively. Owing to the
normalization condition expressed by Eq. (\ref{corre}), $u$ must behave as $r^{\epsilon}$ near the origin with $\mathrm{Re}(\epsilon)>0$ so that only the particular solution $M_{-i\eta,\gamma_{l}}(z)$ with $\gamma_{l}>0$ is allowed. The solution can be written as
\begin{equation}\label{soluregu}
u(z)=A e^{-z/2}z^{1/2+\gamma_{l}}M\left(
1/2+\gamma_{l}+i\eta,1+2\gamma_{l},z \right)
\end{equation}
\noindent where $A$ is a arbitrary constant,
\begin{equation}  \label{vin1}
\alpha_{2}>-1/4 \,,
\end{equation}
\noindent and $M\left( a,b,z \right)$ is the confluent hypergeometric
function (Kummer's function) \cite{ABRAMOWITZ1965} with the asymptotic
behavior for large $\vert z \vert$ and $-\frac{3\pi}{2}<\mathrm{arg}\, z
\leq -\frac{\pi}{2}$ \cite{ABRAMOWITZ1965}
\begin{equation}  \label{asymp}
M\left(a,b,z\right)\simeq \frac{\Gamma(b)}{\Gamma(b-a)}e^{-i\pi a}z^{-a}+
\frac{\Gamma(b)}{\Gamma(a)}e^{z}z^{a-b}\,.
\end{equation}

\subsection{Scattering states}

\label{subsec:2:3}

We can show that for $|z|\gg 1$ and $K\in \mathbb{R}$ the asymptotic
behavior dictated by (\ref{asymp}) implies
\begin{equation}
u(r)\simeq \sin \left( Kr-\frac{l\pi }{2}+\delta _{l}\right)
\label{solassy}
\end{equation}%
\noindent where the relativistic Coulomb phase shift $\delta _{l}=\delta
_{l}\left( \eta \right) $ is given by
\begin{equation}
\delta _{l}=\frac{\pi }{2}\left( l+1/2-\gamma _{l}\right) +\mathrm{arg}%
\,\Gamma \left( 1/2+\gamma _{l}+i\eta \right) .  \label{pshift}
\end{equation}%
\noindent For scattering states, the solution of the DKP equation (\ref%
{dkp11}) has the asymptotic form
\begin{equation}
e^{-i\Lambda }\phi _{1}\left( \vec{r}\right) \simeq e^{iKr\cos \theta
}+f\left( \theta ,\varphi \right) \frac{e^{iKr}}{r}\,,  \label{phiassy}
\end{equation}%
\noindent where the first term represents a plane wave moving along the
direction $\theta =0$ toward the scatterer, and the second term represents a
radially outgoing wave. For spherically symmetric scatterers, both terms
exhibit cylindrical symmetry about the direction of incidence in such a way
that $\phi _{1}$ and $f$ are independent of $\varphi $. The connection
between the forms (\ref{sp}) and (\ref{phiassy}) allows us to write the
scattering amplitude as a partial wave series
\begin{equation}
f\left( \theta \right) =\sum_{l=0}^{\infty }\left( 2l+1\right)
f_{l}P_{l}\left( \cos \theta \right) \,,  \label{serie1}
\end{equation}%
\noindent where $P_{l}$ is the Legendre polynomial of order $l$ and the
partial scattering amplitude is $f_{l}=\left( e^{2i\delta _{l}}-1\right)
/(2iK)$. With the phase shift (\ref{pshift}), up to a logarithmic phase
inherent to the Coulomb field, we find
\begin{equation}
2iKf_{l}=-1+e^{i\pi \left( l+1/2-\gamma _{l}\right) }\frac{\Gamma \left(
1/2+\gamma _{l}+i\eta \right) }{\Gamma \left( 1/2+\gamma _{l}-i\eta \right) }%
\,.  \label{fl}
\end{equation}%
\noindent The series (\ref{serie1}) can be summed only when $\gamma
_{l}=l+1/2$, i.e. when
\begin{equation}
\alpha _{2}=0\,.  \label{c1}
\end{equation}%
\noindent The closed form being \cite{PRC91:034903:2015}
\begin{equation}
f\left( \theta \right) =-\eta \frac{\Gamma \left( 1+i\eta \right) }{\Gamma
\left( 1-i\eta \right) }\frac{e^{-i\eta \ln \sin ^{2}\theta /2}}{2K\sin
^{2}\theta /2}\,,\quad \theta \neq 0\,,  \label{fthe}
\end{equation}%
\noindent which gives the well--known Rutherford scattering formula for the
differential cross section in classical and nonrelativistic quantum
mechanics. This happens because the condition (\ref{c1}) implies that the
scalar boson is effectively under the influence of a Coulomb potential
(without inversely quadratic terms).

\subsection{Bound states}

\label{subsec:2:4}

If $K=i|K|$, the partial scattering amplitude becomes infinite when $%
1/2+\gamma _{l}+i\eta =-n$, where $n=0,1,2,\ldots $, due to the poles of the
gamma function in the numerator of (\ref{fl}), and (\ref{asymp}) implies
that $u$ tends to $r^{1/2+\gamma _{l}+n}e^{-|K|r}$ for large $r$. Therefore,
bound-state solutions are possible only for $\alpha _{1}<0$, i.e when $%
E\gtrless 0$ and $a_{0}^{(1)}\lessgtr 0$ and the spectrum is expressed as
\begin{equation}\label{enerq}
E =-\frac{sgn\left( a_{0}^{(1)}\right) m}{\sqrt{1+\frac{\left(
a_{0}^{(1)}\right) ^{2}}{(n+\gamma _{l}+1/2)^{2}}}}\,.  
\end{equation}%
\noindent Because $M(-n,b,z)$ is proportional to the generalized Laguerre polynomial $L_{n}^{\left(b-1\right) }(z)$ \cite{ABRAMOWITZ1965} and using a pair of integral formulas involving $L_{n}^{\left(b-1\right) }(z)$ \cite{ARFKEN1996} one can finally write $\phi_{1}$ as
\begin{eqnarray}
\phi _{1}(\vec{r}) &=&Ne^{i\Lambda }r^{\gamma _{l}-1/2}\,e^{-\frac{|\alpha
_{1}|r}{2\left( n+\gamma _{l}+1/2\right) }}  \notag \\
&&\times \,L_{n}^{\left( 2\gamma _{l}\right) }\left( \frac{|\alpha _{1}|r}{%
n+\gamma _{l}+1/2}\right) Y_{lm_{l}}(\theta ,\varphi )  \label{normaliza}
\end{eqnarray}%
\noindent with
\begin{equation}\label{N}
N=\sqrt{\frac{|E|}{m}\frac{2^{2\gamma_{l}+1}|K|^{2\gamma_{l}+2}}{(n+\gamma
_{l}+1/2)}\frac{n!}{\Gamma (n+2\gamma _{l}+1) }}\,
\end{equation}%
\noindent where the radial quantum number $n$ is related to the
number of zeros of $\phi_{1}$. In addition to the essential degeneracy mentioned
before, accidental degeneracy exists when $\alpha _{2}=0$. In this case $E$
depends on $n$ and $l$ through the combination $n+l$ in such a way that each
energy is $(n+l+1)^{2}$-fold degenerate.

\section{Conclusions}

\label{sec:3}

In this work, we have addressed the relativistic quantum dynamics of scalar
bosons embedded in a full vector Coulomb interaction in the context of the
DKP formalism. We showed that using the proper form of the full vector
interaction and considering a Coulomb potential (whether attractive or
whe\-ther repulsive) the scattering and bound-state solutions can be obtained
by solving a Whittaker differential equation.

For scattering solutions, we calculated the relativistic Coulomb phase shift
and expressed the scattering amplitude as a partial wave sum for the general
case of vector interactions. The space component of the minimal coupling
contributes only with an $l$-independent phase factor in the partial wave.
The scattering amplitude can be summed only when $\gamma _{l}=l+1/2$ ($%
\alpha _{2}=0$). In this particular case, the closed form leads to the
Rutherford scattering formula for the differential cross section. Incidentally, the
Rutherford scattering formula requires the presence of the space component
of the nonminimal coupling and it is valid even if the interactions are
strong. Furthermore, the Rutherford cross section vanishes in the absence of
the time component of the minimal coupling making the interaction
transparent. Of course, our results are valid in the approximation $\gamma
_{l}\simeq l+1/2$ ($\alpha _{2}\simeq 0$). 

On the other hand, the existence of bound-state solutions requires the
presence of the time component of the minimal coupling $(a_{0}^{(1)}\neq 0)$%
, binding particles (antiparticles) with positive (negative) energies in the
case of an attractive (a repulsive) potential and so the spontaneous pair
production is not a possibility. The usual accidental degeneracy in the
bound-state spectrum only exists when the potential parameters obey the very
same restriction that leads the closed-form amplitude scattering ($\alpha
_{2}=0$, or $\alpha _{2}\simeq 0$ in a approximation scheme). 

We showed that the many ways that a Coulomb potential can couple to bosons
in the DKP formalism make a difference in hadronic process. We believe that
the results reported in the present work are an important contribution for a
better understanding of those phenomenological descriptions by the DKP
formalism. A more detailed study of the interaction between mesons and nucleus can be accomplished by adding a short--range phenomenological potential apart from long--range. A refined calculation of this process should use computational methods to obtain phase--shifts which can be compared with our results obtained in a closed form. Finally, it is worthwhile to mention that the DKP formalism holds a large number of couplings, not only electromagnetic interactions, that enables emulate a mean field theory for describing the hadron interactions. This large number of couplings makes the DKP formalism a great tool for physicists that use a phenomenological description of nuclear processes. Our results can be seen as a first step for future applications in nuclear processes, this is currently under study and will be reported elsewhere.

\begin{acknowledgements}
The authors would like to thank to the referees for useful comments and suggestions. This work was supported in part by means of funds provided by CNPq, Brazil, Grants No. 455719/2014-4 (Universal), No. 304105/2014-7 (PQ) and No. 304743/2015-1 (PQ) and CAPES, Brazil.
\end{acknowledgements}

\appendix
\section{Lorentz Covariance of the DKP equation}
\label{A:1}

Under a Lorentz transformation $x^{\prime}\,^{\mu}=\Lambda^{\mu}\,_{\nu}x^{\nu}$ we have that
\begin{eqnarray}
\psi^{\prime} &=&U(\Lambda)\psi\,,  \label{ap1} \\
U^{-1}\beta^{\mu}U
&=& \Lambda^{\mu}\,_{\nu}\beta^{\nu}  \label{ap2}
\end{eqnarray}%
\noindent For the case of an infinitesimal transformation, we have $\Lambda^{\mu}\,_{\nu}=\delta^{\mu}\,_{\nu}+\Delta\omega^{\mu}\,_{\nu}$ ($\Delta\omega_{\mu\nu}=-\Delta\omega_{\nu\mu}$) with
\begin{equation}\label{ap3}
U=1+\frac{1}{2}\Delta_{\mu\nu}S^{\mu\nu}\,,
\end{equation}
\noindent where $S^{\mu\nu}=\left[ \beta^{\mu},\beta^{\nu} \right]$ and for a finite Lorentz transformation we obtain
\begin{equation}\label{ap4}
U=\mathrm{exp}\left( \frac{1}{2}\Delta\omega_{\mu\nu}S^{\mu\nu} \right)\,.
\end{equation}

\subsection{Scalar sector}
\label{A:1:1}

To select the physical component of the DKP field for the scalar sector (spin-$0$ sector), we define the operators \cite{UMEZAWA1956}
\begin{equation}\label{proj1}
P\equiv-\left( \beta^{0}\right)^{2}\left( \beta^{1}\right)^{2}\left( \beta^{2}\right)^{2}\left( \beta^{3}\right)^{2}, \quad P^{\mu}\equiv P\beta^{\mu}
\end{equation}
\noindent which satisfy
\begin{eqnarray}
P^{2} &=& P\,,  \label{pros1} \\
P^{\mu}\beta^{\nu} &=& Pg^{\mu\nu}\,, \label{pros2}\\
PS^{\mu\nu} &=& S^{\mu\nu}P=0\,, \label{pros3}\\
P^{\mu}S^{\nu\lambda} &=& \eta^{\mu\nu}P^{\lambda}-\eta^{\mu\lambda}P^{\nu} \,. \label{pros4}
\end{eqnarray}%
\noindent As it is shown in \cite{UMEZAWA1956}
\begin{eqnarray}
(P\psi)^{\prime} &=& P\psi\,,\label{psies1}\\
(P^{\mu}\psi)^{\prime} &=& \Lambda^{\mu}\,_{\lambda}P^{\lambda}\psi \,,\label{psies2}
\end{eqnarray}
\noindent so that $P\psi$ and $P^{\mu}\psi$ transform as a (pseudo) scalar and a (pseudo) vector under an infinitesimal Lorentz transformation, respectively. Applying $P$ and $P^{\mu}$ to the DKP equation (\ref{dkp}) and combining the results, we get
\begin{equation}\label{kgfree}
\left(\partial_{\mu}\partial^{\mu}+m^{2}\right)(P\psi)=0\,.
\end{equation} 
\noindent This result tell us that all elements of the column matrix $P\psi$ obey the KG equation. Then, acting $P$ upon the spinor DKP $\psi$ selects the scalar sector of DKP theory, making explicitly clear that it describes a spin-$0$ particle. Following this innovative view of the DKP spinor, Ref. \cite{PLA268:165:2000} shows that the redundant components of $\psi$ are projected out, $\psi$ and $P\psi$ are both compatible with gauge invariance.

\subsection{Vector sector}
\label{A:1:2}

Now we discuss the vector sector (spin-$1$ sector) of the DKP theory. Similar to the scalar sector, we can select the physical components of the DKP field for the spin-$1$ sector, so we define the operators
\begin{equation}\label{proj2}
R^{\mu }\equiv\left( \beta ^{1}\right) ^{2}\left( \beta ^{2}\right) ^{2}\left(
\beta ^{3}\right) ^{2}\left( \beta ^{\mu }\beta ^{0}-g^{\mu 0}\right), \quad
R^{\mu \nu }\equiv R^{\mu }\beta ^{\nu }
\end{equation}%
\noindent which satisfy 
\begin{eqnarray}
R^{\mu \nu } &=& -R^{\nu \mu }\,,\label{prov1}\\
R^{\mu \nu }\beta ^{\alpha } &=& R^{\mu }g^{\nu \alpha }-R^{\nu }g^{\mu\alpha }\,,\label{prov2}\\
R^{\mu}S^{\nu\lambda} &=& g^{\mu\nu}R^{\lambda}-g^{\mu\lambda}R^{\nu}\,,\label{prov3}\\
R^{\mu\nu}S^{\alpha\beta} &=& g^{\nu\alpha}R^{\mu\beta}-g^{\mu\alpha}R^{\nu\beta}-g^{\nu\beta}R^{\mu\alpha}+g^{\mu\beta}R^{\nu\alpha}\,.\label{prov4}
\end{eqnarray}
\noindent As it is shown in \cite{UMEZAWA1956}
 \begin{eqnarray}
(R^{\mu}\psi)^{\prime} &=& \Lambda^{\mu}\,_{\lambda}R^{\lambda}\psi\,,\label{psive1}\\
(R^{\mu\nu}\psi)^{\prime} &=& \Lambda^{\mu}\,_{\lambda}\Lambda^{\nu}\,_{\beta}R^{\alpha\beta}\psi \,,\label{psive2}
\end{eqnarray}
\noindent so that $R^{\mu}\psi$ and $R^{\mu\nu}\psi$ transform as (pseudo) vector and (pseudo) tensor quantities under an infinitesimal Lorentz transformation, respectively. Again, applying $R^{\mu}$ and $R^{\mu\nu}$ to the DKP equation (\ref{dkp}) and combining the results, we obtain
\begin{equation}\label{proca1}
\partial_{\nu}U^{\nu\mu}+m^{2}R^{\mu}\psi=0\,,
\end{equation}
\begin{equation}\label{proca2}
U^{\mu\nu}=\partial^{\mu}R^{\nu}\psi-\partial^{\nu}R^{\mu}\psi\,.
\end{equation}
\noindent These results tell us that all elements of the column matrix $R^{\mu}\psi$ obey the Proca equation. So, similar to the scalar sector, this procedure selects the vector sector of the DKP theory, making explicitly clear that it describes a spin-$1$ particle.

\section{Symmetries of the DKP equation}
\label{A:2}

\subsection{Parity}
\label{A:2:1}
The DKP equation is invariant under the parity operation, i.e. when $\vec{r}\rightarrow-\vec{r}$, if $A_{i}^{(1)}$ and $A_{i}^{(2)}$ change sign, whereas $A_{0}^{(1)}$ and $A_{0}^{(2)}$ remain the same. This happens due to the parity operator is 
\begin{equation}\label{paridade}
\mathcal{P}=\mathrm{exp}\left( i\delta_{P} \right)P_{0}\eta^{0}\,,
\end{equation}
\noindent where $\delta_{P}$ is a constant phase and $P_{0}$ changes $\vec{r}$ into $-\vec{r}$. Because this unitary operator anticommutes with $\beta^{i}$ and $\left[ P,\beta^{i} \right]$, they change sign under a parity transformation, whereas $\beta^{0}$ and $\left[ P,\beta^{0} \right]$, which commute with $\eta^{0}$, remain the same. Since $\delta_{P}=0$ or $\delta_{P}=\pi$, the spinor components have definite parities.

\subsection{Charge--conjugation}
\label{A:2:2}
The charge--conjugation operation changes the sign of the minimal interaction potential, i.e. changes the sign of $A_{\mu}^{(1)}$. This can be accomplished by the transformation $\psi\rightarrow\psi_{C}=\mathcal{C}\psi=CK\psi$, where $K$ denotes the complex conjugation and $C$ is a unitary matrix such that $C\beta^{\mu}=-\beta^{\mu}C$. The matrix that satisfies this relation is 
\begin{equation}\label{conjcar}
C=\mathrm{exp}\left( i\delta_{C} \right)\eta^{0}\eta^{1}\,.
\end{equation}
\noindent The phase factor $\mathrm{exp}\left( i\delta_{C} \right)$ is equal to $\pm 1$; therefore $E\rightarrow-E$. Note also that $J^{\mu}\rightarrow-J^{\mu}$, as should be expected for a charge current. Meanwhile $C$ anticommutes with $\left[ P,\beta^{\mu} \right]$ and the charge--conjugation operation entails no change on $A_{\mu}^{(2)}$.

\subsection{Time--reversal}
\label{A:2:3}
The DKP equation is invariant under the time--reversal transformation, i.e. when $t\rightarrow-t$, if $A_{i}^{(1)}$ and $A_{0}^{(2)}$ change sign, whereas $A_{0}^{(1)}$ and $A_{i}^{(2)}$ remain the same. This is because the time--reversal operator is 
\begin{equation}\label{timerev}
\mathcal{T}=\mathrm{exp}\left( i\delta_{T} \right)T_{0}\eta^{0}\,,
\end{equation}
\noindent where $\delta_{T}$ is a constant phase and $T_{0}$ denotes the complex conjugation and changes $t$ into $-t$. Because this unitary operator anticommutes with $\beta^{i}$ and $\left[ P,\beta^{i} \right]$, they change sign under a time--reversal transformation, whereas $\beta^{0}$ and $\left[ P,\beta^{0} \right]$ remain the same. 

\subsection{$\mathcal{P}\mathcal{C}\mathcal{T}$}
\label{A:2:4}

The DKP equation is invariant under the $\mathcal{P}\mathcal{C}\mathcal{T}$ transformation, if both sorts of vector potential, change sign. Our results can be summarized in Table \ref{table1}.

\begin{table}[h]
 \centering
\begin{tabular}{ c c c c c } \hline
 Vector interaction & $\mathcal{P}$ & $\mathcal{C}$ & $\mathcal{T}$ & $\mathcal{P}\mathcal{C}\mathcal{T}$ \\  \hline
$A_{0}^{(1)}$ & + & - & + & - \\  \hline
$A_{i}^{(1)}$ & - & - & - & - \\  \hline
$A_{0}^{(2)}$ & + & + & - & - \\  \hline
$A_{i}^{(2)}$ & - & + & + & - \\  \hline
\end{tabular}
 \caption{\label{table1} Summary of the results for the behavior of the minimal and nonminimal vector interactions under $\mathcal{P}$, $\mathcal{C}$, $\mathcal{T}$ and $\mathcal{P}\mathcal{C}\mathcal{T}$. Here ``$+$'' and ``$-$'' mean ``no change sign'' and ``change sign'', respectively.}
\end{table}

\bibliographystyle{spphys}       % APS-like style for physics
\bibliography{mybibfile_stars2}

\end{document}